\documentclass[letterpaper]{article} % DO NOT CHANGE THIS
\usepackage{aaai2026}  % DO NOT CHANGE THIS
\usepackage{times}  % DO NOT CHANGE THIS
\usepackage{helvet}  % DO NOT CHANGE THIS
\usepackage{courier}  % DO NOT CHANGE THIS
\usepackage[hyphens]{url}  % DO NOT CHANGE THIS
\usepackage{graphicx} % DO NOT CHANGE THIS
\usepackage{natbib}  % DO NOT CHANGE THIS AND DO NOT ADD ANY OPTIONS TO IT
\usepackage{caption} % DO NOT CHANGE THIS AND DO NOT ADD ANY OPTIONS TO IT
\usepackage{algorithm}
\usepackage{algorithmic}

\usepackage{newfloat}
\usepackage{listings}
\DeclareCaptionStyle{ruled}{labelfont=normalfont,labelsep=colon,strut=off} % DO NOT CHANGE THIS
\floatstyle{ruled}
\newfloat{listing}{tb}{lst}{}
\floatname{listing}{Listing}
\title{Hearing More with Less: Multi-Modal Retrieval-and-Selection Augmented Conversational LLM-Based ASR}
\author{
    Bingshen Mu\textsuperscript{\rm 1}, Hexin Liu\textsuperscript{\rm 2}$^{\ast}$, Hongfei Xue\textsuperscript{\rm 1}, Kun Wei\textsuperscript{\rm 1}, Lei Xie\textsuperscript{\rm 1}\thanks{Corresponding Author.}
}
\affiliations{
    \textsuperscript{\rm 1}Audio, Speech and Language Processing Group (ASLP@NPU), School of Computer Science,\\ Northwestern Polytechnical University, Xi’an, China\\
    \textsuperscript{\rm 2}College of Computing and Data Science, Nanyang Technological University, Singapore
}

\usepackage{bibentry}
\usepackage{amssymb, amsmath}
\usepackage{multirow, booktabs, pifont}

\begin{document}

\maketitle

\begin{abstract}
Automatic Speech Recognition (ASR) aims to convert human speech content into corresponding text. In conversational scenarios, effectively utilizing context can enhance its accuracy.
Large Language Models' (LLMs) exceptional long-context understanding and reasoning abilities enable LLM-based ASR (LLM-ASR) to leverage historical context for recognizing conversational speech, which has a high degree of contextual relevance.
However, existing conversational LLM-ASR methods use a fixed number of preceding utterances or the entire conversation history as context, resulting in significant ASR confusion and computational costs due to massive irrelevant and redundant information.
This paper proposes a multi-modal retrieval-and-selection method named \textbf{MARS} that augments conversational LLM-ASR by enabling it to retrieve and select the most relevant acoustic and textual historical context for the current utterance. 
Specifically, multi-modal retrieval obtains a set of candidate historical contexts, each exhibiting high acoustic or textual similarity to the current utterance.
Multi-modal selection calculates the acoustic and textual similarities for each retrieved candidate historical context and, by employing our proposed near-ideal ranking method to consider both similarities, selects the best historical context.
Evaluations on the Interspeech 2025 Multilingual Conversational Speech Language Model Challenge dataset show that the LLM-ASR, when trained on only 1.5K hours of data and equipped with the MARS, outperforms the state-of-the-art top-ranking system trained on 179K hours of data.
\end{abstract}


\section{Introduction}
Automatic speech recognition (ASR) aims to convert human speech content into corresponding text.
With the development of applications such as speech dialogue systems and meeting transcription, conversational ASR is becoming increasingly crucial.
Typical ASR scenarios involve close-talk single speaker speech, mainly from telephone or audiobook recordings. 
However, conversational speech reflects the complexity of human communication, including diverse speaking styles such as specific speaker pronunciations and vocabulary preferences, paralinguistic phenomena like fillers, stutters, and repairs, as well as a high degree of contextual relevance.
Previous studies indicate that incorporating speech and text modality context from preceding utterances can significantly improve conversational ASR performance~\cite{wei2024conversational, cui2023towards, shon2024generative, gong2023longfnt, hou2022bring, zhang2025mamba}.

\begin{figure}[t]
\centering
\includegraphics[width=1.0\columnwidth]{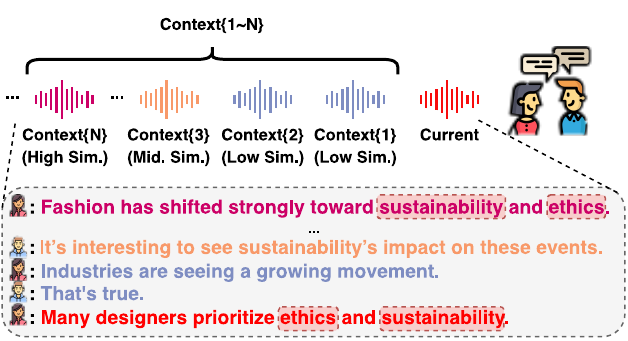}
\caption{Similarity and content of current utterances and historical contexts in conversational speech. The closer the historical context is to red, the more similar it is to the current utterance. The purple ``Context\{N\}" represents the most relevant historical context, and  ``Context\{1$\sim$N\} refers to the preceding N contexts.}
\label{fig1}
\end{figure}
Large language models (LLMs) have demonstrated exceptional long-context understanding and reasoning abilities~\cite{touvron2023llama1, touvron2023llama2, bai2023qwen}, making them promising components for conversational ASR.
% The exceptionally long-context understanding and reasoning abilities of Large Language Models (LLMs)~\cite{achiam2023gpt, touvron2023llama1, touvron2023llama2, bai2023qwen} make them promising components for conversational ASR.
Recent studies combining LLMs with conversational ASR involve extending LLM-based ASR (LLM-ASR) models~\cite{geng2024unveiling, bai2024seed, mu2025hdmole} by additionally inputting two types of context: a fixed number of preceding utterances~\cite{peng2025bi, li2025seewo} and the entire conversation history~\cite{bai2024seed, ding2025kimi, xu2025qwen2}.
The former methods assume that the most relevant historical context to the current utterance appears in the preceding few utterances, overlooking the fact that these utterances contain a large number of fillers and other semantically insignificant contexts, which can severely affect the ASR of the current utterance. 
% Moreover, as shown in Figure~\ref{fig1}, since the position of the most relevant historical context is uncertain, a fixed number of preceding utterances may not include the most relevant historical context to the current utterance.
Figure~\ref{fig1} shows that the position of the most relevant historical context for the current utterance is not fixed; it may be located in earlier conversation history, beyond a fixed number of preceding utterances.
The latter methods use the entire conversation history as context. Although this provides richer context, it inevitably leads to redundant information that interferes with the ASR and incurs significant computational costs.
These limitations highlight the need for a more efficient and contextually effective strategy for leveraging conversational history.
Consequently, the research question can be summarized as follows: \textit{How can we retrieve and select the most relevant historical context for the current utterance to augment conversational LLM-ASR performance}.

% Retrieval Augmented Generation (RAG) has garnered widespread attention for its ability to enhance accurate, up-to-date, verifiable, and domain-specific text generation by integrating large-scale external knowledge retrieval systems with LLMs~\cite{arslan2024survey, fan2024survey, wu2024retrieval, mei2025survey}.
Retrieval Augmented Generation (RAG) provides a potential solution to conversational LLM-ASR because it can enhance accurate, up-to-date, verifiable, and domain-specific text generation by integrating large-scale external knowledge retrieval systems with LLMs~\cite{arslan2024survey, fan2024survey, wu2024retrieval, mei2025survey, abootorabi2025ask, ni2025towards}.
However, RAG exhibits limited adaptability in conversational ASR.
RAG focuses on generating new content based on retrieved text, whereas ASR aims to map speech to text. Thus, the two objectives are fundamentally different.
Furthermore, the vast amount of content retrieved by RAG can lead to information overload in ASR, drowning out the speech that needs to be recognized.
Nevertheless, the RAG design philosophy can be applied to retrieve and select the best historical context from conversational speech to assist in recognizing the current utterance.

This paper proposes a multi-modal retrieval-and-selection method named \textbf{MARS} for enhancing conversational LLM-ASR. 
% Unlike existing approaches that rely on either a fixed number of preceding utterances or the entire conversation history, MARS retrieves and selects a historical context of comparable length to the current utterance, removing redundancy while retaining the most relevant information from the whole history.
MARS retrieves and selects a historical context of comparable length to the current utterance, removing redundancy while retaining the most relevant information from the whole history.
Specifically, multi-modal retrieval obtains a set of candidate historical contexts, each exhibiting high acoustic or textual similarity to the current utterance.
Subsequently, multi-modal selection calculates the acoustic and textual similarities for each retrieved candidate historical context, and our proposed near-ideal ranking method considers both similarities and selects the best historical context. 
By simultaneously inputting the hypothesis from the best historical context and the current utterance speech embedding with its hypothesis into LLM-ASR, optimal ASR performance can be achieved.
Experiments on the Interspeech 2025 Multilingual Conversational Speech Language Model (MLC-SLM) Challenge dataset~\cite{mu2025summary} validate the effectiveness of MARS. It achieves a significantly lower Mixed Error Rate (MER) using only 1.5K hours of training data, outperforming the top-ranking method in the challenge, which employs a specially designed LLM-ASR trained on 179K hours of data.
It is worth noting that MARS sets a new state-of-the-art on the MLC-SLM dataset.
\section{Related Work}
\subsection{Conversational LLM-ASR}
Previous LLM-ASR research primarily focused on isolated utterances due to real-world conversational ASR data scarcity.
The release of the MLC-SLM dataset has gradually invigorated research in conversational LLM-ASR.
The TEA-ASLP system enhances Ideal-LLM~\cite{xue2024ideal} with language identification and multilingual Mixture-of-Experts (MoE), achieving optimal performance after pretraining on 179K hours of multilingual data and fine-tuning on 1.5K hours of MLC-SLM data. However, their attempt to incorporate historical and future context during the fine-tuning phase yielded no benefits~\cite{xue2025teaaslp}.
The Seewo system explores the upper bound of potential benefits from historical context by utilizing the ground-truth content of the two preceding utterances~\cite{li2025seewo}.
\citeauthor{peng2025bi} investigates the impact of using both historical and future context on conversational LLM-ASR. It proposes a character-level context masking strategy during training, where portions of the context are randomly removed to enhance robustness and better simulate potential faulty transcriptions that may occur during inference.
\subsection{RAG in Speech LLMs}
RAG leverages external knowledge to make LLMs generate more accurate and contextually relevant content, with widespread applications in the speech modality.
SEAL~\cite{sun2025seal} explores using a shared embedding space for speech-to-text retrieval, focusing on evaluating the quality of learned embeddings through proxy tasks.
WavRAG~\cite{chen2025wavrag} leverages the speech encoding abilities of Qwen-Audio~\cite{chu2023qwen} to generate semantically rich speech embeddings for retrieval.
\citeauthor{feng2025enhancing} proposes an RAG framework for speech-to-speech dialogue models, which can retrieve textual knowledge with input speech.
\citeauthor{pusateri2025retrieval} uses a RAG-like technique to correct ASR entity name errors by querying a vector database with error-prone ASR hypotheses and feeding the retrieved entities and hypotheses to an adapted LLM for error correction.
SpeechT-RAG~\cite{zhang2025speecht} improves depression detection by using speech timing features as the basis for RAG.
\begin{figure*}[t]
\centering
\includegraphics[width=1.0\textwidth]{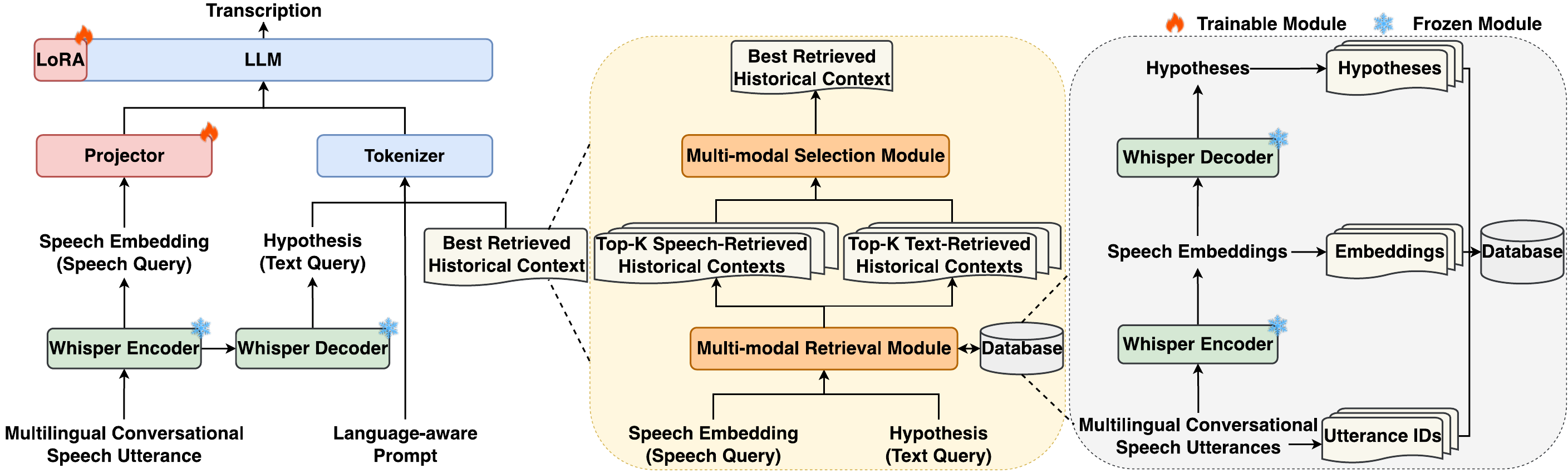} 
\caption{Overview of MARS. \textbf{\textit{Left}}: a language-aware prompt, the best retrieved historical context, the current utterance speech embedding, and its hypothesis are used as inputs to the LLM for jointly training the projector and LoRA parameters to predict the transcription; \textbf{\textit{Middle}}: the multi-modal retrieval module retrieves historical contexts from the database based on speech and text queries, while the multi-modal selection module determines the best historical context from the retrieved ones; \textbf{\textit{Right}}: the database is constructed using Whisper, storing a triplet for each utterance in the conversation.}
% \caption{Overview of MARS. \textbf{\textit{(a)}} the database is constructed using Whisper, storing a triplet for each utterance in the conversation; \textbf{\textit{(b)}} a language-aware prompt, the best retrieved historical context, the current utterance speech embedding, and its hypothesis are used as inputs to the LLM for jointly training the projector and LoRA parameters to predict the transcription; \textbf{\textit{(c)}} the multi-modal retrieval module retrieves historical contexts from the database based on speech and text queries, while the multi-modal selection module determines the best historical context from the retrieved ones.}
\label{fig2}
\end{figure*}
\section{MARS}
\subsection{Overview}
The objective of MARS is to determine the best historical context for the current utterance through the multi-modal retrieval-and-selection method, thereby improving ASR accuracy.
The overall framework of MARS is illustrated in Figure~\ref{fig2}.
The database is constructed using fully fine-tuned Whisper. It stores a triplet for each utterance in conversational speech, consisting of the utterance's ID, speech embedding, and hypothesis.
In MARS, speech embeddings and hypotheses serve as speech and text queries, respectively. The multi-modal retrieval module uses these queries to retrieve the Top-K most similar historical contexts from the database.
Even after retrieval, multiple historical contexts still cause ASR prediction confusion due to information redundancy, and excessively long historical contexts will consume significant computational costs. The multi-modal selection module determines the best historical context from the retrieved contexts. 
Then, a language-aware prompt, the best retrieved historical context, the current utterance speech embedding, and its hypothesis input to the LLM for jointly training the projector and low-rank adaptation (LoRA) parameters~\cite{hu2022lora} to predict the transcription.
\subsection{Multi-modal Retrieval}
\begin{figure}[t]
\centering
\includegraphics[width=1.0\columnwidth]{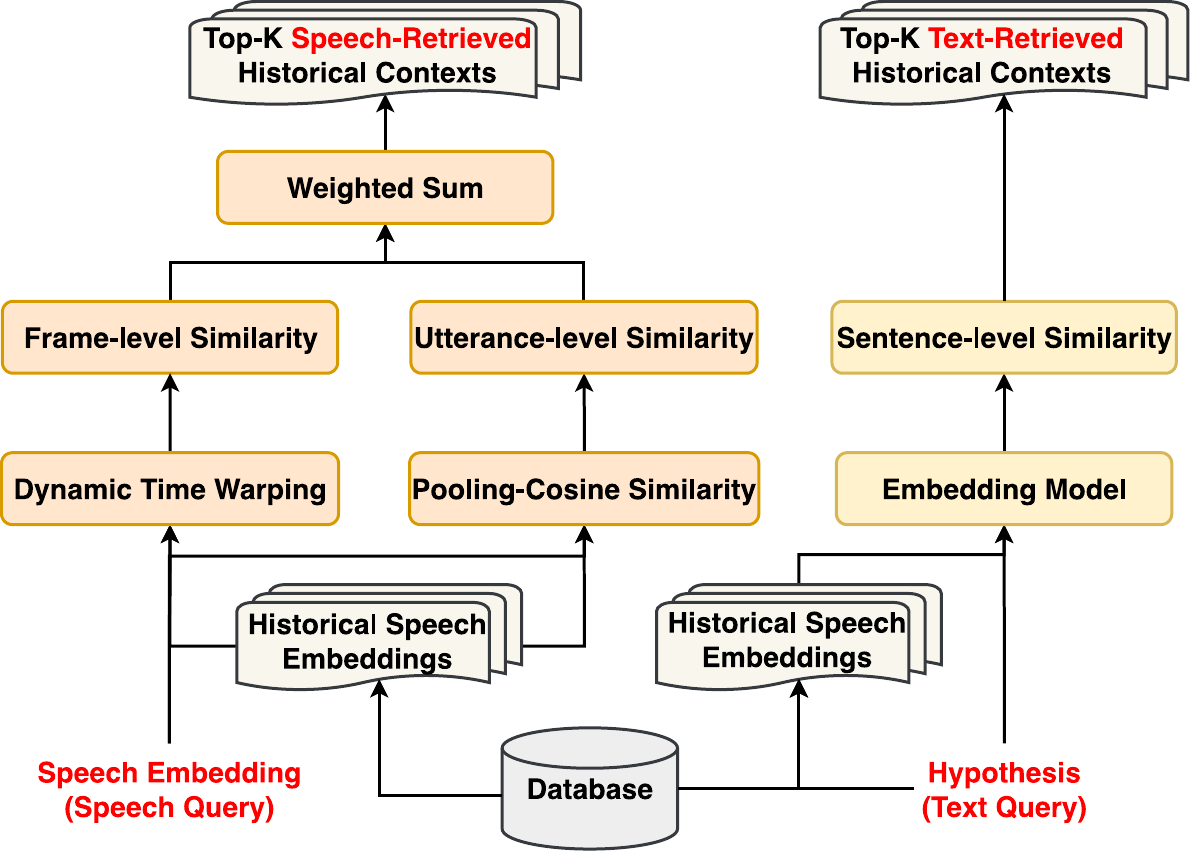}
\caption{The details of the multi-modal retrieval module.}
\label{fig3}
\end{figure}

In conversational ASR, both speech and text modalities carry crucial contextual information.
Relying on a single modality alone cannot comprehensively retrieve historical contexts similar to the current utterance in a conversation~\cite{wei2022leveraging}. 
In contrast, multi-modal retrieval measures the similarity of historical contexts from both speech and text perspectives, providing historical contexts for the current utterance from both pronunciation and semantic aspects.
Historical contexts retrieved by speech modality can reduce ASR errors caused by pronunciation variations, while historical contexts retrieved by text modality can reduce ASR errors caused by word ambiguities.
Figure~\ref{fig3} illustrates the detailed pipeline of the multi-modal retrieval module.

In speech modality retrieval, we use the Dynamic Time Warping (DTW) to calculate the frame-level acoustic similarity between the current and historical speech embeddings.
DTW calculates the matching path between two speech embeddings to determine their minimum cumulative distance.
However, traditional DTW is highly computationally complex. Retrieving a large number of historical speech embeddings for the current utterance can take significant time.
Therefore, we use FastDTW~\cite{salvador2007toward}, which significantly reduces computational complexity while maintaining high accuracy.
Additionally, we calculate utterance-level acoustic similarity by pooling the current and historical speech embeddings and computing their cosine similarity.
After weighted summing the frame-level and utterance-level similarities, we obtain the speech retrieval similarities of the historical speech embeddings relative to the current utterance.
We select the Top-K historical contexts with the highest speech retrieval similarities as the Top-K speech-retrieved historical contexts for the current utterance.

In text modality retrieval, we use the embedding model to calculate the sentence-level semantic similarities between the current and historical utterance hypotheses, serving as the text retrieval similarities. We then select the Top-K historical utterance hypotheses with the highest text retrieval similarities as the Top-K text-retrieved historical contexts for the current utterance.
\subsection{Multi-modal Selection}
After multi-modal retrieval, we observe that using Top-K speech-retrieved or text-retrieved historical contexts, either individually or in combination, degrades ASR performance.
However, when we select the best historical context from either speech-retrieved or text-retrieved, the ASR performance improves.
Therefore, we present a multi-modal selection module to determine the best historical context from the Top-K speech-retrieved and text-retrieved historical contexts.
This method not only enhances ASR performance but also mitigates the issue of increased computational cost caused by excessively long contexts.
\begin{figure}[t]
\centering
\includegraphics[width=1.0\columnwidth]{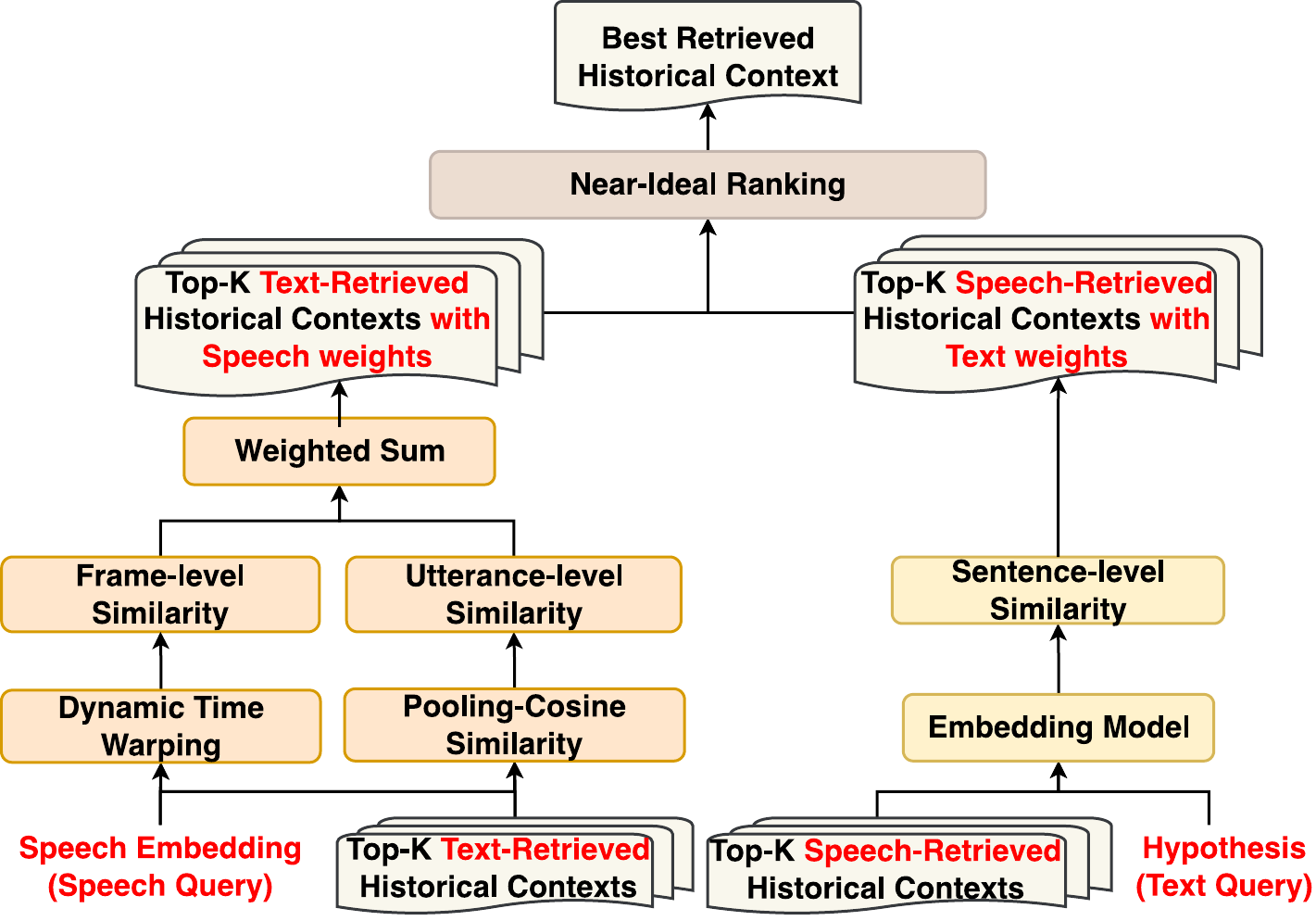}
\caption{The details of the multi-modal selection module.}
\label{fig4}
\end{figure}

Figure~\ref{fig4} illustrates the detailed pipeline of the multi-modal selection module.
We first calculate the speech and text retrieval similarities for all retrieved historical contexts.
For the Top-K speech-retrieved historical contexts, each has a speech retrieval similarity but no text retrieval similarity. 
Therefore, we calculate the text retrieval similarity for each context, ultimately ensuring that the Top-K speech-retrieved historical contexts have both speech and text retrieval similarities. 
At the same time, we calculate the speech retrieval similarities for the Top-K text-retrieved historical contexts.
After obtaining the speech and text retrieval similarities for all retrieved historical contexts, the challenge lies in combining these two similarities to determine the best retrieved historical context.
Since the speech and text retrieval similarities are computed using different methods, these similarities do not have the same dimensions and cannot be directly converted through an algorithm.
Furthermore, selecting the best retrieved historical context requires simultaneously considering both speech and text retrieval similarities. Therefore, directly summing these two similarities and then ranking to choose the maximum sum is not a reasonable approach.

To address the above challenges, we propose an approach called near-ideal ranking, which simultaneously considers both speech and text retrieval similarities to determine the best retrieved historical context.
Assume there are a total of $2K$ retrieved historical contexts. 
For the $i$-th historical context, its speech and text retrieval similarity are denoted as $sw_i$ and $tw_i$, respectively.
We first construct a decision matrix containing both speech and text retrieval similarities and normalize them to eliminate the differences.
The normalization process can be denoted as:
\begin{equation}
    \begin{aligned}
        & sr_i = \frac{sw_i}{\sqrt{ {\textstyle \sum_{j=1}^{2K}} sw_j^2} }, \\
        & tr_i = \frac{tw_i}{\sqrt{ {\textstyle \sum_{j=1}^{2K}} tw_j^2} }, \\
    \end{aligned}
\end{equation}
where $sr_i$ and $tr_i$ are the $i$-th historical context with normalized speech and text retrieval similarities.
Then, we define the ideal as a virtual historical context where both speech and text retrieval similarities are optimal:
\begin{equation}
    \begin{aligned}
        & sa^+ = max\left \{ sr_1, sr_2, ..., sr_{2K} \right \},  \\
        & ta^+ = max\left \{ tr_1, tr_2, ..., tr_{2K} \right \}, \\
    \end{aligned}
\end{equation}
where $sa^+$ and $ta^+$ are the speech and text retrieval similarities of the ideal.
The negative ideal is a virtual historical context where both speech and text retrieval similarities are the worst:
\begin{equation}
    \begin{aligned}
        & sa^- = min\left \{ sr_1, sr_2, ..., sr_{2K} \right \},  \\
        & ta^- = min\left \{ tr_1, tr_2, ..., tr_{2K} \right \}, \\
    \end{aligned}
\end{equation}
where $sa^-$ and $ta^-$ are the speech and text retrieval similarities of the negative ideal.
Next, we calculate the Euclidean distance between the retrieved historical contexts and both the ideal and negative ideal:
\begin{equation}
    \begin{aligned}
        & d_i^+ = \sqrt{(sr_i-sa^+)^2+(tr_i-ta^+)^2},  \\
        & d_i^- = \sqrt{(sr_i-sa^-)^2+(tr_i-ta^-)^2}, \\
    \end{aligned}
\end{equation}
where $d_i^+$ and $d_i^-$ are the Euclidean distance between $i$-th retrieved historical context both the ideal and negative ideal.
Finally, we compute the relative closeness of each retrieved historical context, defined as its Euclidean distance to the negative ideal divided by the sum of its Euclidean distances to both the ideal and negative ideal:
\begin{equation}
    c_i = \frac{d_i^-}{d_i^+ + d_i^-},
\end{equation}
where $c_i$ is the relative closeness of the $i$-th retrieved historical context.
A relative closeness closer to 1 indicates that the retrieved historical context is better.
The historical context with the maximum relative closeness is the best retrieved historical context for the current utterance.
\subsection{Adaptive Contextual Decoding}
During the training of the conversational LLM-ASR, we randomly decide whether to use the best retrieved historical context.
This training strategy of randomly masking the best retrieved historical context can enhance the generalization capability of the conversational LLM-ASR, preventing it from over-relying on historical context and neglecting the current utterance itself.
\begin{table*}[t]
\centering
\begin{tabular}{lcccccccccc}
\toprule
\multirow{2}{*}{Language} & \multicolumn{2}{c}{Vanilla Whisper} & \multicolumn{2}{c}{Fine-tuned Whisper} & \multicolumn{2}{c}{Qwen2-Audio} & \multicolumn{2}{c}{TEA-ASLP} & \multicolumn{2}{c}{MARS} \\ \cmidrule{2-11} 
                          & Dev              & Test             & Dev                   & Test                 & Dev            & Test           & Dev           & Test         & Dev         & Test        \\ \midrule
English-American          & 14.14            & 13.79            & 10.77                 & 9.27                 & 12.58          & 12.27          & 9.12          & 9.13         & 9.37        & 8.31        \\
English-American          & 11.72            & 10.97            & 7.22                  & 6.97                 & 13.77          & 12.89          & 6.23          & 7.41         & 5.99        & 5.75        \\
English-British           & 10.08            & 8.45             & 7.35                  & 5.98                 & 12.32          & 10.33          & 6.36          & 5.81         & 6.14        & 4.88        \\
English-Filipino          & 9.20             & 8.47             & 6.66                  & 7.05                 & 14.94          & 13.75          & 6.18          & 7.61         & 5.61        & 6.20        \\
English-Indian            & 13.96            & 8.89             & 12.61                 & 8.26                 & 21.66          & 13.79          & 11.66         & 7.64         & 10.17       & 7.19        \\
French                    & 28.14            & 29.94            & 14.31                 & 16.78                & 70.61          & 75.13          & 13.74         & 17.33        & 10.98       & 13.10       \\
German                    & 20.72            & 18.36            & 18.99                 & 15.47                & 91.52          & 81.10          & 16.89         & 15.10        & 14.62       & 13.14       \\
Italian                   & 17.92            & 19.97            & 12.25                 & 13.73                & 70.95          & 79.07          & 11.78         & 12.55        & 10.37       & 11.86       \\
Japanese                  & 21.64            & 18.81            & 14.23                 & 13.76                & 21.02          & 18.27          & 13.61         & 9.88         & 11.93       & 11.18       \\
Korean                    & 13.80            & 10.72            & 8.69                  & 7.53                 & 37.85          & 29.40          & 8.64          & 6.28         & 7.72        & 6.57        \\
Portuguese                & 21.23            & 23.47            & 17.20                 & 16.29                & 115.86         & 128.08         & 19.61         & 17.73        & 13.64       & 13.46       \\
Russian                   & 17.67            & 16.99            & 12.73                 & 11.51                & 62.86          & 60.44          & 12.55         & 13.62        & 11.45       & 10.45       \\
Spanish                   & 12.27            & 14.09            & 8.51                  & 7.87                 & 91.85          & 105.47         & 8.37          & 9.51         & 7.21        & 6.91        \\
Thai                      & 14.49            & 22.92            & 12.76                 & 8.12                 & 101.03         & 159.81         & 8.45          & 7.09         & 6.46        & 5.64        \\
Vietnamese                & 27.16            & 19.69            & 11.18                 & 7.39                 & 103.28         & 74.87          & 11.45         & 6.64         & 9.64        & 6.52        \\ \midrule
Avg.                      & 16.82            & 17.33            & 11.87                 & 10.15                & 51.90          & 53.47          & 10.62         & 9.60         & \textbf{8.97}        & \textbf{8.35}        \\ \bottomrule
\end{tabular}
\caption{The WER (\%) $\downarrow$ and CER (\%) $\downarrow$ results for each language, as well as the average MER (\%) $\downarrow$ results across all languages for our proposed MARS and other comparative baseline methods on the MLC-SLM dev and test datasets.}
\label{table1}
\end{table*}

Furthermore, the conversational LLM-ASR trained with this strategy can adapt to various decoding strategies:
\begin{itemize}
    \item \textbf{Direct decoding:} Each utterance is decoded independently in the conversational LLM-ASR, without reliance on any historical context.
    \item \textbf{MARS decoding:} Each utterance is decoded in the conversational LLM-ASR by combining it with the best retrieved historical context, which is determined through the multi-modal retrieval-and-selection method.
    \item \textbf{Two-pass decoding:} In the first pass decoding, a preliminary hypothesis for each utterance is obtained through direct decoding. Subsequently, a new database is constructed to store utterances and their preliminary hypotheses. In the second pass decoding, the final predicted transcription for each utterance is obtained through the MARS decoding, which determines the best retrieved historical context from the newly constructed database.
\end{itemize}
\section{Experimental Setup}
\subsection{Dataset}
We conduct our experiments on the MLC-SLM dataset, which originates from the recently held Interspeech 2025 Multilingual Conversational Speech Language Model Challenge~\cite{mu2025summary}.
The dataset comprises 11 languages: English, French, German, Italian, Portuguese, Spanish, Japanese, Korean, Russian, Thai, and Vietnamese.
The English subset comprises approximately 500 hours of recordings from various regions, including British, American, Australian, Indian, and Philippine English. Other languages contribute around 100 hours each, resulting in a total of approximately 1500 hours of multilingual conversational speech data.
Each recording consists of a multi-turn, natural, and fluent conversational speech of around 20 minutes between two speakers on a randomly assigned topic, including celebrities, dreams, education, emotion, fashion, food, games, the Internet, movies, shopping, travel, etc, recorded using devices such as iPhones in a quiet indoor environment.

\subsection{Implementation Details}
We construct the database using the Whisper-large-v3 fully fine-tuned on the MLC-SLM dataset.
For the MARS, we use the fine-tuned Whisper encoder, and the input hypotheses are pre-generated through inference by the fine-tuned Whisper model.
The projector consists of two linear layers connected by a ReLU activation function.
Moreover, we employ Qwen2.5-7B-Instruct~\cite{team2024qwen2} as the LLM, with LoRA configured with a rank of 64, an alpha value of 256, and a dropout rate of 0.05.
Seven modules in each LLM layer, including q\_proj, k\_proj, v\_proj, o\_proj, up\_proj, gate\_proj, and down\_proj, are subject to the LoRA.
In speech modality retrieval, the frame-level and utterance-level similarities are summed with weights of 0.5 each. In text modality retrieval, we use the Qwen3-Embedding-0.6B~\cite{zhang2025qwen3} to calculate the similarities.
\begin{table}[t]
\centering
\begin{tabular}{lcc}
\toprule
Model                       & Context Type                           & Dev   \\ \midrule
\multirow{3}{*}{Bi-context} & None                                   & 14.87 \\
                            & Context\{1$\sim$2\} \& Future\{1\}           & 13.56 \\
                            & GT: Context\{1$\sim$2\} \& Future\{1\} & 13.16 \\ \midrule
\multirow{2}{*}{Seewo}      & None                                   & 15.48 \\
                            & GT: Context\{1$\sim$2\}                     & 14.30 \\ \midrule
\multirow{2}{*}{MARS}       & None                                   & 12.75 \\
                            & Best Retrieved                         & \textbf{8.96}  \\ \bottomrule
\end{tabular}
\caption{The MER (\%) $\downarrow$ results of using various types of context on the MLC-SLM dev dataset. ``GT" refers to using the ground-truth transcription as context.}
\label{table2}
\end{table}
All language-aware prompts convey the same meaning: ``Please transcribe the speech into text".
The specific language prompt chosen corresponds to the language identification of the utterance.
The multi-modal retrieval module generates the Top-3 speech-retrieved and text-retrieved historical contexts separately.
When training MARS, we randomly mask the best retrieved historical context with a 50\% probability.
The MARS is trained on 6 NVIDIA A800 GPUs, with a maximum batch size accommodating 30 seconds of speech and a gradient accumulation of 6. 
The Adam optimizer is used with a warm-up scheduler that adjusts the learning rate, peaking at 0.0001 after 200 steps.
Each model is trained for 3 epochs, and all checkpoints are averaged for the final inference.
During inference, the LLM generates transcriptions without employing sampling methods, with the beam size, temperature, repetition penalty, and length penalty all set to 1.
\subsection{Evaluation Metrics}
Following common evaluation standards for multilingual ASR, for languages with character-based writing systems and no clear word boundaries—including Japanese, Korean, and Thai—we use Character Error Rate (CER) to measure ASR performance. 
For all other languages, we use Word Error Rate (WER).
Additionally, we use the Mixed Error Rate (MER) to measure the average ASR error rate across 11 languages.
To ensure a fair comparison with the solutions from the MLC-SLM challenge, we use the MeetEval toolkit~\cite{von2023meeteval} to calculate all ASR error rates.
\section{Experimental Results}
\subsection{Main Results}
Table~\ref{table1} presents the WER and CER results for each language, as well as the average MER results across all languages for MARS and other comparative methods, including: 1) Vanilla Whisper-large-v3~\cite{radford2023robust}, which demonstrates excellent ASR performance across a wide range of languages; 2) Fully Fine-tuned Whisper-large-v3, which fine-tuned on the MLC-SLM training dataset; 3) Qwen2-Audio~\cite{chu2024qwen2}, an speech LLM adaptable to various speech tasks and showing strong ASR performance; and 4) TEA-ASLP~\cite{xue2025teaaslp}, an LLM-ASR model trained on 179K hours of multilingual ASR data including the MLC-SLM training dataset, achieved the state-of-the-art performance in the MLC-SLM test dataset.
The vanilla Whisper-large-v3 demonstrates good ASR performance on the MLC-SLM dev and test datasets, and fine-tuning further improves its performance.
Qwen2-Audio performs relatively well on the English subsets, but its performance on other languages is inferior due to insufficient multilingual training data.
TEA-ASLP demonstrates excellent ASR performance across all languages after large-scale training. With the support of a multilingual LLM, it outperforms the fine-tuned Whisper model.
MARS, using only the 1.5K hours MLC-SLM training dataset, outperforms TEA-ASLP in the majority of languages by effectively leveraging relevant historical context in conversational speech.
MARS demonstrates the significant potential of retrieving and selecting suitable historical context to augment conversational ASR, highlighting remarkable data utilization that achieves high accuracy with significantly less training data.

Table~\ref{table2} illustrates the comparison results of MARS with other methods that augment conversational LLM-ASR using context.
Bi-context~\cite{peng2025bi} reports the results of using two preceding contexts and one future context, while Seewo~\cite{li2025seewo} reports the results of using two preceding contexts. They also evaluate ground-truth transcriptions as context to explore the upper bound of their methods.
Even with ground-truth transcription as context, the benefits are limited, indicating that the immediate preceding utterances still contain irrelevant and redundant information.
Furthermore, the relative gains they achieved from utilizing context are inferior to those of MARS, even when using ground-truth transcriptions.
The results further underscore the necessity of MARS in retrieving and selecting the best historical context.
\begin{table}[t]
\centering
\begin{tabular}{lcc}
\toprule
Model                  & Dev   & Test  \\ \midrule
LLM-ASR                & 12.75 & 11.04 \\
\hspace{1em}+ Hyp.      & 11.15 & 9.89  \\
\hspace{2em}+ Speech Retrieval      & 10.24 & 9.41  \\
\hspace{2em}+ Text Retrieval        & 10.33 & 9.23  \\
\hspace{2em}+ Multi-modal Retrieval & 11.49 & 9.34  \\
\hspace{3em}+ Multi-modal Selection                 & 9.77  & 8.96  \\
\hspace{4em}+ Two-pass Decoding      & \textbf{8.97}  & \textbf{8.35}  \\ \bottomrule
\end{tabular}
\caption{The ablation study MER (\%) $\downarrow$ results of removing each component of MARS on the MLC-SLM test dataset.}
\label{table3}
\end{table}
\begin{table}[t]
\centering
\begin{tabular}{lccc}
\toprule
Retrieval Type                         & Selection Type & Dev   & Test  \\ \midrule
\multirow{5}{*}{None}                  & Context\{1\}         & 11.01 & 9.74  \\
                                       & Context\{1$\sim$2\}         & 11.86 & 9.90  \\
                                       & Context\{1$\sim$3\}         & 11.75 & 10.42 \\
                                       & Context\{1$\sim$4\}         & 11.72 & 11.98 \\
                                       & Context\{1$\sim$5\}         & 12.51 & 13.49 \\ \midrule
\multirow{3}{*}{Speech}      & Top-1          & 10.24 & 9.41  \\
                                       & Top-2          & 11.53 & 9.65  \\
                                       & Top-3          & 11.23 & 9.92  \\ \midrule
\multirow{3}{*}{Text}        & Top-1          & 10.33 & 9.23  \\
                                       & Top-2          & 10.72 & 9.41  \\
                                       & Top-3          & 10.90 & 9.68  \\ \midrule
\multirow{5}{*}{Multi-modal} & 2$\times$Top-1        & 11.49 & 9.34  \\
                                       & 2$\times$Top-2        & 10.11 & 10.04 \\
                                       & 2$\times$Top-3        & 10.56 & 11.24 \\
                                       & Sum \& Top-1    & 10.19 & 9.18  \\
                                       & Multi-modal    & \textbf{9.77}  & \textbf{8.96}  \\ \bottomrule
\end{tabular}
\caption{The ablation study MER (\%) $\downarrow$ results of various retrieval and selection types of MARS on the MLC-SLM test dataset. ``Sum" means the sum of retrieval similarities.}
\label{table4}
\end{table}
\subsection{Ablation Study}
The ablation study in Table~\ref{table3} demonstrates the effectiveness of each component of MARS.
Incorporating the ASR hypothesis of the current utterance into LLM-ASR can improve performance.
% In the experiments on speech or text retrieval of historical contexts, we select the historical context with the highest similarity.
After incorporating historical contexts with the highest speech or text retrieval similarity to the current utterance, ASR performance further improves. We also observe that text-retrieved historical contexts are superior to those speech-retrieved.
The performance of multi-modal retrieval is not as effective as that of speech or text retrieval because we select the most similar speech and text retrieval contexts, and the redundant information from the two historical contexts interferes with the ASR process.
Therefore, it is essential to select the best historical context from the Top-K speech-retrieved and text-retrieved historical contexts generated by multi-modal retrieval.
Applying multi-modal selection after multi-modal retrieval, specifically by using the near-ideal ranking method to choose the best historical context, can effectively improve ASR performance.
Finally, randomly masking historical contexts during training and utilizing two-pass decoding to leverage more accurate historical contexts for the current utterance's ASR yields the best results for MARS.
\begin{figure}[t]
\centering
\includegraphics[width=1.0\columnwidth]{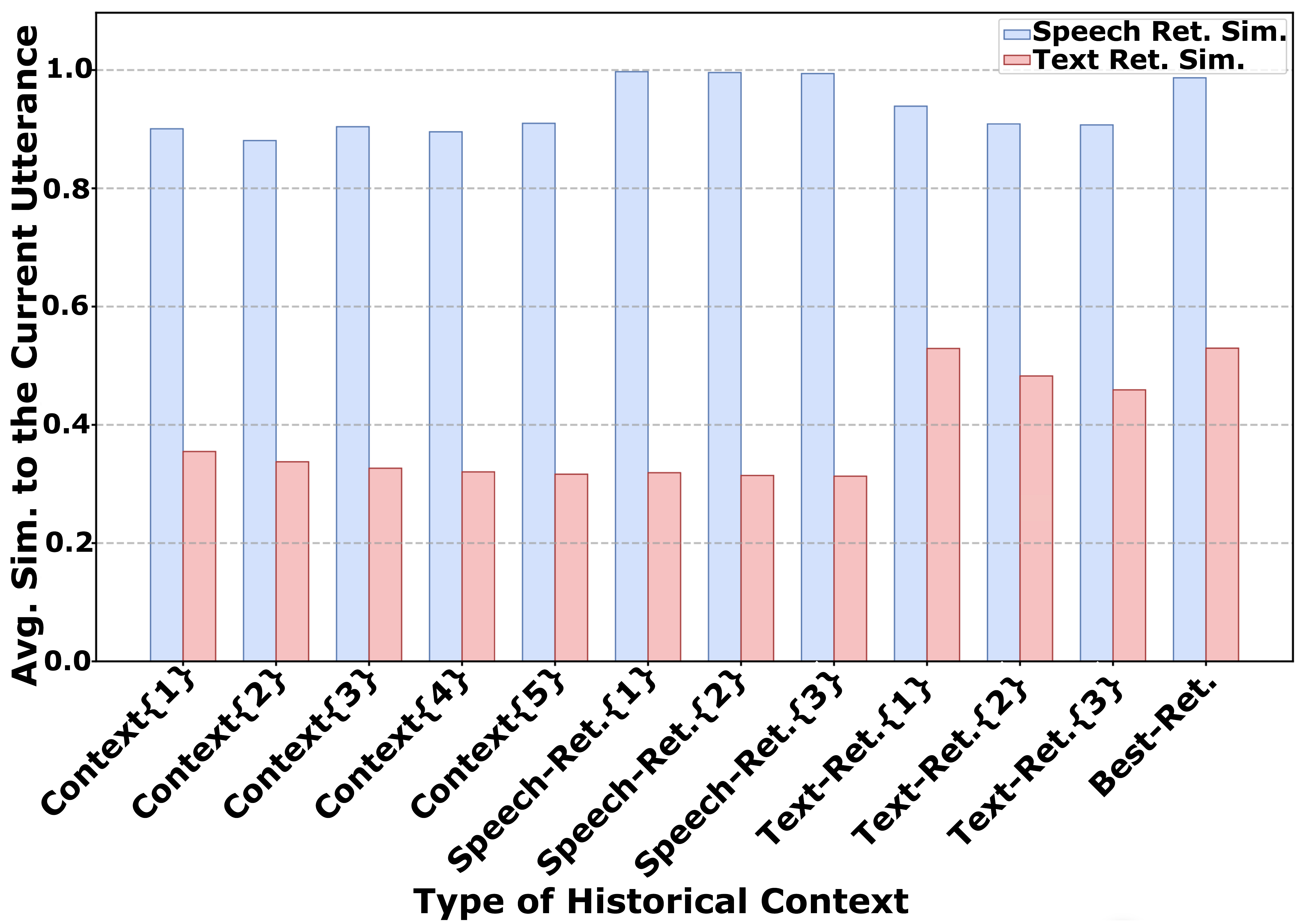}
\caption{Average speech and text retrieval similarity between different types of historical contexts and the current utterance in the MLC-SLM test dataset.}
\label{fig5}
\end{figure}
\begin{figure}[t]
\centering
\includegraphics[width=1.0\columnwidth]{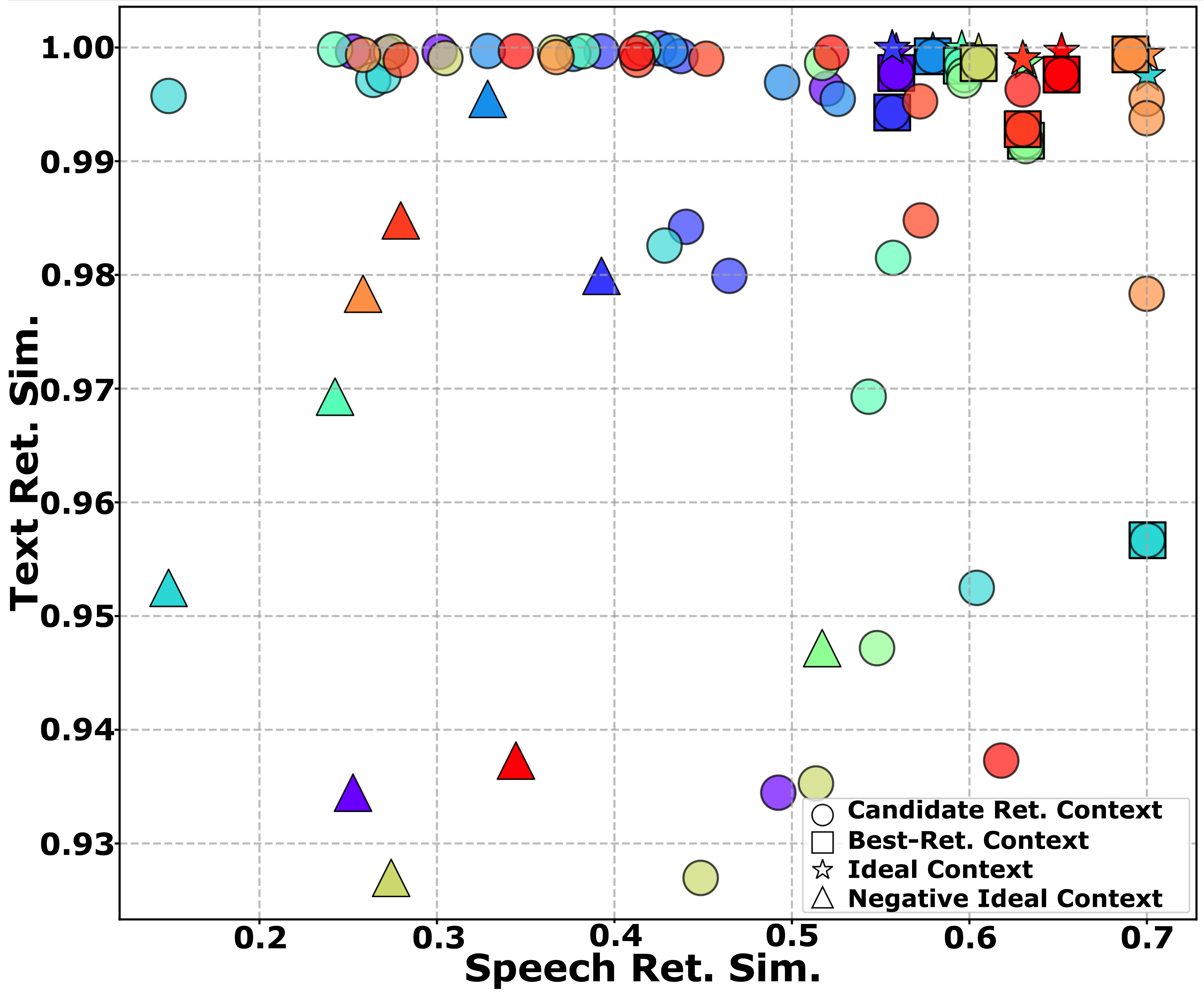}
\caption{Visualization of our proposed near-ideal ranking method on 10 randomly selected utterances from the MLC-SLM test dataset.}
\label{fig6}
\end{figure}

Additionally, we further conduct detailed ablation experiments on multi-modal retrieval and multi-modal selection in Table~\ref{table4}.
Under the condition of the same number of historical contexts, using speech or text retrieval historical contexts outperforms using a fixed number of preceding contexts, which demonstrates the necessity of retrieving historical contexts.
Moreover, we find that as the number of historical contexts increased, ASR performance degraded significantly, indicating that excessive historical context leads to information redundancy, which is detrimental to the recognition of the current utterance.
After multi-modal retrieval, a total of 2K historical contexts are obtained. 
To fully leverage the potential of retrieved historical contexts, we need to select the best one from these.
Compared to multi-modal selection, simply summing the speech and text retrieval similarities of each retrieved historical context and selecting the one with the highest total similarity performs worse, which validates the effectiveness of multi-modal selection.

\subsection{Visualization}
Figure~\ref{fig5} illustrates the average speech and text retrieval similarity between different types of historical contexts, including preceding, speech retrieval, and text retrieval, and the current utterance in the MLC-SLM test dataset.
We observe that the preceding historical contexts have lower average speech and text retrieval similarity compared to retrieved historical contexts.
Historical contexts retrieved by speech or text possess considerably higher speech or text retrieval similarity.
The best-retrieved historical context obtained through multi-modal selection exhibits speech and text retrieval similarities that are close to those of the speech-retrieved and text-retrieved historical contexts, demonstrating the effectiveness of the multi-modal selection in choosing the best historical context.
Moreover, the different similarity calculation methods result in significant numerical differences between speech and text retrieval similarities. 
Directly summing both to obtain the best historical context is not appropriate, highlighting the advantages of the near-ideal ranking method.

Figure 6 visualizes the near-ideal ranking method across 10 randomly selected utterances from the MLC-SLM test dataset.
We observe that the best historical context selected by this method has speech and text retrieval similarities that closely align with the ideal assumption, where both values are maximized, and are far from the negative ideal assumption, where both values are minimized.
The near-ideal ranking method not only avoids the issue of chaotic ASR results caused by information redundancy from using multiple historical contexts, but also significantly improves ASR performance and reduces computational cost by utilizing only the best historical context.
\section{Conclusion}
In this paper, we propose a multi-modal retrieval-and-selection method named \textbf{MARS} for conversational LLM-ASR. 
Multi-modal selection obtains a set of candidate historical contexts, each exhibiting high acoustic or textual similarity to the current utterance.
Subsequently, multi-modal selection calculates the acoustic and textual similarities for each retrieved candidate historical context, and our proposed near-ideal ranking method considers both similarities and selects the best historical context. 
Evaluations on the Interspeech 2025 MLC-SLM Challenge dataset validate the effectiveness of MARS, which receives and selects the most relevant historical context for the current utterance to augment conversational LLM-ASR. Furthermore, the results show that the LLM-ASR, when trained on only 1.5K hours of data and equipped with the MARS, outperforms the state-of-the-art top-ranking system trained on 179K hours of data.
\bibliography{aaai2026}
\end{document}